\documentclass[preprint,showpacs,aps]{revtex4}
\usepackage{graphicx}
\usepackage{graphics}
\usepackage{amsmath}
\usepackage{amssymb}

\begin{document}

\title{Statistical mechanics of semiflexible polymers}
\author{Semjon Stepanow }
\affiliation{Martin-Luther-Universit\"{a}t Halle, Fachbereich Physik, D-06099\\
Halle, Germany}
\date{\today }

\begin{abstract}
We present the statistical-mechanical theory of semiflexible polymers based
on the connection between the Kratky-Porod model and the quantum rigid
rotator in an external homogeneous field, and treatment of the latter using
the quantum mechanical propagator method. The expressions and relations
existing for flexible polymers can be generalized to semiflexible ones, if
one replaces the Fourier-Laplace transform of the end-to-end polymer
distance, $1/(k^{2}/3+p)$, through the matrix $\tilde{P}(k,p)=(I+ikDM)^{-1}D$%
, where $D$ and $M$ are related to the spectrum of the quantum rigid
rotator, and considers an appropriate matrix element of the expression under
consideration. The present work provides also the framework to study
polymers in external fields, and problems including the tangents of
semiflexible polymers. We study the structure factor of the polymer, the
transversal fluctuations of a free end of the polymer with fixed tangent of
another end, and the localization of a semiflexible polymer onto an
interface. We obtain the partition function of a semiflexible polymer in
half space with Dirichlet boundary condition in terms of the end-to-end
distribution function of \ the free semiflexible polymer, study the
behaviour of a semiflexible polymer in the vicinity of a surface, and \
adsorption onto a surface.
\end{abstract}

\maketitle

\section{ Introduction}

\label{intro}

Polymers with contour length $L$ much larger than the persistence length $%
l_{p}$, which is the correlation length for the tangent-tangent correlation
function along the polymer and is quantitative measure of the polymer
stiffness, are flexible and are described by using the tools of quantum
mechanics and quantum field theory \cite{edwards65}-\cite{doi-edwards-book}.
If the chain length decreases, the chain stiffness becomes an important
factor. Many polymer molecules have internal stiffness and cannot be
modelled by the model of flexible polymers developed by Edwards \cite%
{edwards65}.

The standard coarse-graining model of a wormlike or a semiflexible polymer
was proposed by Kratky and Porod \cite{kratky/porod49}. A few first moments
of $G(r,N)$ were computed in \cite{hermans/ullman52}-\cite{saitoetal67}. The
literature on the earlier work on semiflexible polymers can be found in the
book by Yamakawa \cite{yamakawa}. For recent work see \cite{warneretal85}-%
\cite{samuel-sinha02} (and the references therein). Despite a considerable
interest and immense efforts in last decades there is no theory of
semiflexible polymers providing a general tool for treating problems
including semiflexible polymers.

In this article we present the theory of semiflexible polymers based on the
relation between the Kratky-Porod model and the quantum rigid rotator in an
external field \cite{descloizeaux},\cite{saitoetal67},\cite{zinn-justin},
and treatment of the latter in the framework of the quantum mechanical
propagator method \cite{feynman-hibbs}. Although the most polymer quantities
are defined through the positions of the monomers $\mathbf{r}(s)$, ($0\leq
s\leq N$) (the end-to-end distribution function, the scattering function,
the isotropic monomer-monomer interactions, etc.), and on the contrary the
quantum rigid rotator is formulated in terms of tangents $\mathbf{t}%
(s)=\partial \mathbf{r}(s)/\partial s$, we have shown that the relations for
flexible polymers (polymers in external fields, polymers with
self-interactions, etc.) can be generalized to semiflexible polymer,
replacing the Fourier-Laplace transform of the end-to-end distribution
function of the flexible polymer, $1/(k^{2}/3+p)$, by the infinite order
matrix $\tilde{P}(k,p)=(I+ikDM)^{-1}D$ with matrices $D$ and $M$ related to
the spectrum of the rigid rotator, and considering an appropriate matrix
element of the matrix expression. The quantity $\tilde{P}(k,p)$ plays the
key role in the theory similar to the bare propagator in common quantum
field theories. The end-to-end distribution function is simply the matrix
element $\left\langle 0,0|\tilde{P}(k,p)|0,0\right\rangle $, the scattering
function of the polymer is the inverse Laplace transform of $G(k,p)/p^{2}$
multiplied by $2/N$, the partition function of the stretched polymer is $%
Z(f,N)=G(k=-iF/k_{B}T,N)$ etc. The elimination of summations over the
magnetic quantum number in intermediate states enables one to carry out the
calculations of above quantities with the infinite order square matrix $%
\tilde{P}^{s}(k,p)$. The present theory provides also the framework to study
semiflexible polymers in external fields and with self-interactions, and
problems including the tangents of polymer configurations.

The article is organized as follows. Sec.\ref{formal} introduces to the
Green's function formalism of the quantum mechanical rigid rotator. Sec.\ref%
{end-to-end} derives the exact expression for the Fourier-Laplace transform
of the end-to-end distribution function, and establishes its important
properties. Sec.\ref{s-fact} presents results of the computation of the
scattering function of a semiflexible polymer. Sec.\ref{ads-if} considers
the localization of semiflexible polymers in a weak symmetric potential
corresponding to adsorption onto an interface. Sec.\ref{dirichlet} considers
the behaviour of the polymer in the presence of a surface. Sec.\ref{ads-surf}
treats the adsorption onto a surface. Sec.\ref{self-inter} introduces to the
description of semiflexible polymers with self-interactions.

\section{Formalism}

\label{formal}

The Fourier transform of the distribution function of the end-to-end polymer
distance of the Kratky-Porod model \cite{kratky/porod49} $G(\mathbf{k}%
,L)=\int d^{3}r\exp (-i\mathbf{k(r-r_{0})})G(\mathbf{r}-\mathbf{r}_{0},L)$
is expressed by the path integral as follows
\begin{equation}
G(\mathbf{k},L)=\int D\mathbf{t}(s)\prod\limits_{s}\delta (\mathbf{t}%
(s)^{2}-1)\exp (-i\mathbf{k}\int_{0}^{L}ds\mathbf{t}(s)-\frac{l_{p}}{2}%
\int_{0}^{L}ds(\frac{\partial \mathbf{t}(s)}{\partial s})^{2}),  \label{w1}
\end{equation}%
where $l_{p}$ is the persistence length, and $\mathbf{t}(s)=\partial \mathbf{%
r}(s)/\partial s$ is the tangent vector at the point $s$ along the contour
length of the polymer. The 2nd term in the exponential is associated with
the bending energy. The product over $s$ in Eq.(\ref{w1}) takes into account
that the polymer chain is locally inextensible. For a polymer which
interacts with an external potential and for polymer with monomer-monomer
interactions the terms $-\int_{0}^{L}dsV(\mathbf{r}(s))$ and $-\frac{1}{2}%
\int_{0}^{L}ds_{2}\int_{0}^{L}ds_{1}U(\mathbf{r}(s_{2})-\mathbf{r}(s_{1}))$
should be added in the exponential of Eq.(\ref{w1}), respectively.

The path integral (\ref{w1}) (without the term depending on $\mathbf{k}$)
corresponds to the diffusion of a particle on unit sphere, $\left\vert
\mathbf{t}(s)\right\vert =1$ \cite{descloizeaux}, \cite{warneretal85}, and
is also equivalent to the Euclidean quantum rigid rotator \cite{descloizeaux}%
,\cite{zinn-justin}. The Green's function of the quantum rigid rotator or
that for diffusion of a particle on unit sphere obeys the following equation%
\begin{equation}
\frac{\partial }{\partial L}P_{0}(\theta ,\varphi ,L;\theta _{0},\varphi
_{0},0)-\frac{1}{2l_{p}}\nabla _{\theta ,\varphi }^{2}P_{0}=\delta (L)\delta
(\Omega -\Omega _{0}),  \label{w2}
\end{equation}%
where $\Omega $ is the spherical angle characterized by angles $\theta $,
and $\varphi $, and $\delta (\Omega -\Omega _{0})$ is a two dimensional
delta function having the property $\int d\Omega \delta (\Omega -\Omega
_{0})=1$. Henceforth, instead of the contour length $L$ we will use the
number of segments $N=L/l_{p}$. All lengths will be measured in units of the
persistence length $l_{p}$. The quantity $P(\theta ,\varphi ,N;\theta
_{0},\varphi _{0},0)$ is the Fourier transform of the end-to-end polymer
distance with fixed tangents of both ends. The Fourier transform of the
distribution function of the end-to-end polymer distance is obtained from $%
P(\Omega ,N;\Omega _{0},0)$ by integrating the latter over $\Omega $ and $%
\Omega _{0}$%
\begin{equation}
G(k,N)=\frac{1}{4\pi }\int d\Omega \int d\Omega _{0}P(\Omega ,N;\Omega
_{0},0).  \label{w6}
\end{equation}%
Notice that in the quantum mechanical counterpart of the problem $N$
corresponds to the imaginary time $it$. The bare Green's function $%
P_{0}(\theta ,\varphi ,N;\theta _{0},\varphi _{0},0)$ associated with (\ref%
{w2}) reads
\begin{equation}
P_{0}(\theta ,\varphi ,N;\theta _{0},\varphi _{0},0)=\sum_{l,m}\exp (-\frac{%
l(l+1)N}{2})Y_{lm}(\theta ,\varphi )Y_{lm}^{\ast }(\theta _{0},\varphi _{0}),
\label{w3}
\end{equation}%
where $Y_{lm}(\theta ,\varphi )$ are the spherical harmonics, and $l$ and $m$
are the quantum numbers of the angular momentum. For a given $l$, $m$ takes
the values $-l$, $-l+1$, $...$, $l$. The quantity $P_{0}(\theta ,\varphi
,N;\theta _{0},\varphi _{0},0)$ corresponds to Eq.(\ref{w1}) with $\mathbf{k}%
=0$ and with the following boundary conditions for the path $\mathbf{t}(s)$ (%
$0\leq s\leq N$): $\mathbf{t}(N)\equiv (\theta ,\varphi )$, and $\mathbf{t}%
(0)\equiv (\theta _{0},\varphi _{0})$.

We now will consider the Green's function $P(\theta ,\varphi ,N;\theta
_{0},\varphi _{0},0)$ associated with Eq.(\ref{w1}). The differential
equation for $P$ is%
\begin{equation}
\frac{\partial }{\partial N}P(\theta ,\varphi ,N;\theta _{0},\varphi _{0},0)-%
\frac{1}{2}\nabla _{\theta ,\varphi }^{2}P+U(\Omega )P=\delta (N)\delta
(\Omega -\Omega _{0}),  \label{w4}
\end{equation}%
where $U(\mathbf{kt}_{\Omega })=i\mathbf{k\mathbf{t}_{\Omega }}$ is the
potential energy of the rigid rotator in an external field $i\mathbf{k}$,
where $\mathbf{k}$ is measured in units of $l_{p}^{-1}$. As it is well-known
from Quantum Mechanics \cite{feynman-hibbs} the differential equation (\ref%
{w4}) can be rewritten as an integral equation as follows%
\begin{equation}
P(\Omega ,N;\Omega _{0},0)=P_{0}(\Omega ,N;\Omega _{0},0)-\int_{0}^{N}ds\int
d\Omega ^{\prime }P_{0}(\Omega ,N;\Omega ^{\prime },s)U(\mathbf{kt}_{\Omega
^{\prime }})P(\Omega ^{\prime },s;\Omega _{0},0).  \label{w5}
\end{equation}%
As we already mentioned above Eqs.(\ref{w3}-\ref{w5}) describes also the
Euclidean rigid quantum rotator in an external field. The iteration of Eq.(%
\ref{w5}) generates the perturbation expansion of $P(\Omega ,N;\Omega
_{0},0) $ in powers of the potential $U(\mathbf{kt}_{\Omega })$, and can be
symbolically written as%
\begin{equation}
P=P_{0}-P_{0}UP_{0}+P_{0}UP_{0}UP_{0}-...=P_{0}-P_{0}UP.  \label{ieq}
\end{equation}%
The coefficient in front of ($k^{2})^{n}$ of the expansion of $G(k,p)$ and
consequently of $P$ in powers of $k^{2}$, multiplied by the factor $%
(-1)^{n}\Gamma (2n+2)$ is the Laplace transform of the moment $<r^{2n}>$ of
the end-to-end distribution function. Thus, Eq.(\ref{ieq}) gives the moment
expansion of the end-to-end distribution function. The integral equation (%
\ref{w5}) is nothing but the Dyson equation. The description of semiflexible
polymers based on (\ref{w5}-\ref{ieq}) is a variant of the application of
the methods of quantum field theory to problems of statistical mechanics
\cite{zinn-justin},\cite{descloizeaux} (and citations therein).

\section{\textbf{The end-to-end distribution function and its moments}}

\label{end-to-end}

\subsection{General consideration}

\label{g-cons}

We now will consider the computation of the end-to-end distribution function
of a semiflexible polymer using the propagator method. The matrix elements
of the external potential $U(\Omega )$ over the spherical functions are
given by
\begin{equation}
\left\langle l^{\prime },m^{\prime }|U|l,m\right\rangle =ik\int d\Omega
Y_{l^{\prime }m^{\prime }}^{\ast }(\theta ,\varphi )(\mathbf{nt_{\Omega })}%
Y_{lm}(\theta ,\varphi )\equiv ikM_{l^{\prime },m^{\prime };l,m}(\theta
_{1},\varphi _{1}),  \label{w7}
\end{equation}%
where the unit vector $\mathbf{t}$ is characterized by the angles $\theta $
and $\varphi $, and $\mathbf{n}=\mathbf{k}/k$ by the angles $\theta _{1}$
and $\varphi _{1}$. Due to the convolution character of the expression (\ref%
{w5}) with respect to the integration over the contour length ($P_{0}(\Omega
,N;\Omega ^{\prime },s)$ depends on the difference $N-s$), the Laplace
transform of $P(\Omega ,N;\Omega _{0},0)$ in Eq.(\ref{w5}) with respect to $%
N $ permits to get rid of integrations over the contour length. Thus, in the
following we will consider the Laplace transform of $G(k,N)$ with respect to
$N$. Using the spectral expansion of $P_{0}$ according to Eq.(\ref{w3}) in
the perturbation expansion of $P(\Omega ,N;\Omega _{0},0)$ which is given by
Eq.(\ref{ieq}) enables one to sum the moment expansion of $G(k,p)$ in powers
of $k$ as%
\begin{equation}
G(k,p)=\left\langle 0,0|\tilde{P}(\mathbf{k},p)|0,0\right\rangle ,
\label{w8}
\end{equation}%
where%
\begin{equation}
\tilde{P}(\mathbf{k},p)=(I+ikDM)^{-1}D,  \label{w9}
\end{equation}%
and $D$ is infinite order matrix defined by%
\begin{equation}
D_{l,l^{\prime }}=\frac{1}{\frac{1}{2}l(l+1)+p}\delta _{l,l^{\prime }}
\label{w10}
\end{equation}%
with $l$, $l^{\prime }=0$, $1$, $...$. Eq.(\ref{w9}) can be also derived via
direct solution of the Dyson equation (\ref{ieq}). The zeros at the brackets
in Eq.(\ref{w8}) means $l=0$ and $m=0$. Notice that Eqs.(\ref{w8}-\ref{w9})
have to be understood as quantum mechanical expectation values, so that
summations over the quantum numbers $l$ and $m$ of the angular momentum in
the intermediate states of the series of $\tilde{P}(\mathbf{k},p)$ are
implied.

We now will show that summations over the magnetic quantum number $m$ in
intermediate states in Eq.(\ref{w8}) can be eliminated. This can be shown
using the fact that $G(k,p)$ is function of $\mathbf{k}^{2}$, which
legitimates us to choose the $z$-axes of the coordinate system along $%
\mathbf{k}$ in computing the matrix elements $\left\langle l^{\prime
},m^{\prime }|U|l,m\right\rangle $. In this case the scalar product $\mathbf{%
nt}$ in Eq.(\ref{w7}) becomes simply $\cos \theta $, with the consequence
that the matrix elements $M_{l^{\prime },m^{\prime };l,m}$ become diagonal
with respect to indices $m$ and $m^{\prime }$, so that the magnetic number
will be zero throughout the products of matrices in the series of $\tilde{P}%
(k,p)$. As a result Eqs.(\ref{w8}-\ref{w9}) simplify to%
\begin{equation}
G(k,p)=\left\langle 0|\tilde{P}^{s}(k,p)|0\right\rangle   \label{w8a}
\end{equation}%
with
\begin{equation}
\tilde{P}^{s}(k,p)=(I+ikDM^{s})^{-1}D,  \label{w9a}
\end{equation}%
where the square matrix $M^{s}$ is defined by%
\begin{equation}
M_{l,l^{\prime }}^{s}=w_{l}\delta _{l,l^{\prime }+1}+w_{l+1}\delta
_{l+1,l^{\prime }},  \label{w11}
\end{equation}%
and $w_{l}=\sqrt{l^{2}/(4l^{2}-1)}$. Summations over the intermediate states
in Eqs.(\ref{w8a}-\ref{w9a}) occur over the eigenvalues of the angular
momentum $l=0,1,...$, so that according to Eqs.(\ref{w8a}-\ref{w9a}) the
calculation of $G(k,p)$ reduces to the computation of the matrix element of
an infinite order \textit{square} matrix \cite{SS02}. Eqs.(\ref{w8}-\ref{w9}%
) and (\ref{w8a}-\ref{w9a}), which are more general as those derived in \cite%
{SS02}, enable one to compute the end-to-end distribution function with
fixed tangents, and to study polymers in external fields and with
self-interactions. The validity of Eqs.(\ref{w8a}-\ref{w9a}) can be proved
in more general way using the relation%
\begin{equation}
\left\langle l_{1},m_{1}|\tilde{P}(\mathbf{k},p)|l_{2},m_{2}\right\rangle
=\sum\limits_{m^{\prime }}D_{m^{\prime },m_{1}}^{l_{1}\ast }(\alpha ,\beta
,\gamma )\left\langle l_{1}|\tilde{P}^{s}(k,p)|l_{2}\right\rangle
D_{m^{\prime },m_{2}}^{l_{2}}(\alpha ,\beta ,\gamma ),  \label{wign}
\end{equation}%
where $D_{m,m^{\prime }}^{l}(\alpha ,\beta ,\gamma )$ is the Wigner $D$%
-function \cite{edmonds57}, $D_{m,m^{\prime }}^{l}(\alpha ,\beta ,\gamma
)=e^{-im\alpha }d_{m,m^{\prime }}^{l}e^{-im^{\prime }\gamma }$, and $\alpha $%
, $\beta $, $\gamma $ are Euler angles chosen such that the $z$-axes of the
transformed coordinate system is directed along the wave vector $\mathbf{k}$%
. In obtaining (\ref{wign}) we have taken into account that the matrix
elements $\left\langle l_{1},m_{1}^{\prime }|\tilde{P}(\mathbf{k}%
,p)|l_{2},m_{2}^{\prime }\right\rangle $ computed in the coordinate system
with the $z$-axes parallel to $\mathbf{k}$ are diagonal with respect to
indices $m_{1}^{\prime }$ and $m_{2}^{\prime }$. Using (\ref{wign}) and the
property of the Wigner $D$-function, $D_{m,0}^{l}(\alpha ,\beta ,\gamma )=%
\sqrt{4\pi /(2l+1)}Y_{lm}^{\ast }(\beta ,\alpha )$, one obtains
\begin{equation}
\left\langle l,m|\tilde{P}(\mathbf{k},p)|0,0\right\rangle =\sqrt{4\pi /(2l+1)%
}Y_{lm}^{\ast }(\theta _{1},\varphi _{1})\left\langle l|\tilde{P}%
^{s}(k,p)|0\right\rangle .  \label{mel1}
\end{equation}%
Eqs.(\ref{wign}-\ref{mel1}) are valid as well for matrix elements of $%
(DM)^{n}$, which are terms in the series of $\tilde{P}(\mathbf{k},p)$ in
powers of $k^{2}$. Notice that summations over intermediate states on the
left side occur over $l$ and $m$, while summations on the right side are
only over $l$, as it is already clear from notations. To establish Eqs.(\ref%
{w8a}-\ref{w9a}) using (\ref{mel1}) we consider the expression%
\begin{equation}
\left\langle 0,0|(DM)^{n_{1}+n_{2}}|0,0\right\rangle =\sum_{l,m}\left\langle
0,0|(DM)^{n_{1}}|l,m\right\rangle \left\langle
l,m|(DM)^{n_{2}}|0,0\right\rangle .  \label{w11c}
\end{equation}%
The application of (\ref{mel1}) to both off-diagonal matrix elements in (\ref%
{w11c}) combined with the use of the addition theorem for spherical
functions, $P_{l}(\cos \omega )=4\pi /(2l+1)\sum_{m=-l}^{l}Y_{lm}(\theta
^{\prime },\varphi ^{\prime })Y_{lm}^{\ast }(\theta ,\varphi )$, with $%
\omega $ being the angle between the vectors characterized by spherical
angles $\theta ^{\prime }$, $\varphi ^{\prime }$ and $\theta $, $\varphi $,
which is zero in the case under consideration, shows that the summation over
$m$ in (\ref{w11c}) will be eliminated, and we obtain again Eq.(\ref{w8a}).
Eq.(\ref{wign}) enables one to replace $\tilde{P}(\mathbf{k},p)$ in favor of
$\tilde{P}^{s}(\mathbf{k},p)$ in the expression
\begin{eqnarray}
\sum\limits_{m}\left\langle l_{1},m_{1}|\tilde{P}(\mathbf{k}%
_{1},p)|l,m\right\rangle \left\langle l,m|\tilde{P}(\mathbf{k}%
_{2},p)|l_{2},m_{2}\right\rangle  &=&\sum\limits_{m_{1}^{\prime
},m_{2}^{\prime }}D_{m_{1}^{\prime },m_{1}}^{l_{1}\ast }(\alpha _{1},\beta
_{1},\gamma _{1})\left\langle l_{1}|\tilde{P}^{s}(k_{1},p)|l\right\rangle
\notag \\
&&\times \sum\limits_{m}D_{m_{1}^{\prime },m}^{l}(\alpha _{1},\beta
_{1},\gamma _{1})D_{m_{2}^{\prime },m}^{l\ast }(\alpha _{2},\beta
_{2},\gamma _{2})  \notag \\
&&\times \left\langle l|\tilde{P}^{s}(k_{2},p)|l_{2}\right\rangle
D_{m_{2}^{\prime },m_{2}}^{l_{2}}(\alpha _{2},\beta _{2},\gamma _{2}),
\label{m}
\end{eqnarray}%
where the Euler angles $\alpha _{i}$, $\beta _{i}$, $\gamma _{i}$ define the
coordinate system with the $z$-axes parallel to the wave vector $\mathbf{k}%
_{i}$. The limits in summations over the magnet quantum numbers in (\ref{m})
are determined by the corresponding quantum number of the angular momentum.
Eq.(\ref{m}) can be generalized in a straightforward way for matrix elements
of products of arbitrary number of propagators $\tilde{P}(\mathbf{k}_{i},p)$%
. The sum over $m$ in (\ref{m}) can be carried out using the addition
formula of the Wigner $D$-function%
\begin{equation*}
\sum\limits_{m^{\prime \prime }}D_{m,m^{\prime \prime }}^{l}(\alpha
_{1},\beta _{1},\gamma _{1})D_{m^{\prime \prime },m^{\prime }}^{l}(\alpha
_{2},\beta _{2},\gamma _{2})=D_{m,m^{\prime }}^{l}(\alpha ,\beta ,\gamma ),
\end{equation*}%
where $\alpha $, $\beta $, $\gamma $ are the Euler angles for the resulting
coordinate transformation $S\rightarrow S_{1}\rightarrow S_{2}$, to give
\begin{equation}
D_{m,m^{\prime }}^{l}(\alpha ^{\prime },\beta ,\gamma ^{\prime }).
\label{sum-m}
\end{equation}%
The angles $\alpha ^{\prime }$ and $\gamma ^{\prime }$ correspond to the
coordinate transformation $S\rightarrow S_{1}\rightarrow S_{2}^{\prime }$
with $\alpha _{2}^{\prime }=-\alpha _{2}$, $\beta _{2}^{\prime }=\beta _{2}$%
, $\gamma _{2}^{\prime }=-\gamma _{2}$. Expressions similar to (\ref{m})
appear in studies of semiflexible polymers in external fields (adsorption)
and with self-interactions, where in the perturbation expansions of the
quantities such as partition function, "propagators" $\tilde{P}(\mathbf{k}%
_{i},p)$ with different wave vectors $\mathbf{k}_{i}$ appear. The
consideration of the expression in the middle line of (\ref{m}) containing
the sum over $m$ in one dimension, where the Euler angles take the values $%
\alpha _{i}=\gamma _{i}=0$, $\beta _{i}=0$, $\pi $ which means that in this
case the wave vectors $\mathbf{k}_{i}$ can be only parallel or antiparallel.
In this case the expression (\ref{sum-m}) reduces to $D_{m,m^{\prime
}}^{l}(0,0,0)=\delta _{m,m^{\prime }}$ or $D_{m,m^{\prime }}^{l}(0,\pi
,0)=(-1)^{l+m}\delta _{m,-m^{\prime }}$. Therefore, the magnet quantum
numbers in the intermediate states of expectation values of expressions like
(\ref{m}) over the ground state will be zero, so that the factors in the
intermediate states (\ref{sum-m}) become simply $(\pm 1)^{l}$. The sign
minus applies, if the neighbor wave vectors are antiparallel.

In explicit computations based on Eqs.(\ref{w8a}-\ref{w9a}) one should
truncate the infinite square matrix $\tilde{P}^{s}(k,p)$ by a finite matrix
of order $n$. The expression for $G(k,p)$ obtained in this way is a rational
function being an infinite series in powers of $k^{2}$, i.e. it contains all
moments of the end-to-end distribution function, and describes the first $%
2n-2$ moments exactly. In context of eigenstates of the rigid rotator, the
truncation at order $n$ takes into account the eigenstates with quantum
number up to the value $l=n-1$. The truncation of $\tilde{P}^{s}(k,p)$ by $4$
order matrix, which is the consequence of corresponding truncation of the
matrices $D$ and $M^{s}$, yields the Fourier-Laplace transform of the
end-to-end distribution function as follows%
\begin{equation}
G_{6}(k,p)=\frac{1}{p}\frac{1+\frac{4\,k^{2}}{15\,\left( 1+p\right) \,\left(
3+p\right) }+\frac{9\,k^{2}}{35\,\left( 3+p\right) \,\left( 6+p\right) }}{%
\,1+\frac{k^{2}}{3\,p\,\left( 1+p\right) }+\frac{4\,k^{2}}{15\,\left(
1+p\right) \,\left( 3+p\right) }+\frac{9\,k^{2}}{35\,\left( 3+p\right)
\,\left( 6+p\right) }+\frac{3\,k^{4}}{35\,p\,\left( 1+p\right) \,\left(
3+p\right) \,\left( 6+p\right) }}.  \label{gkp6}
\end{equation}%
$G_{6}(k,p)$ is the infinite series in powers of $k^{2}$, and describes
exactly the first $6$ moments of the end-to-end-distribution function. The
moments of the end-to-end distance are obtained from (\ref{w8a}) as
\begin{equation}
\left\langle R^{2n-2}\right\rangle =\Gamma (2n)\mathcal{L}%
_{p}^{-1}\left\langle 0|(DM^{s})^{2n-2}|0\right\rangle ,  \label{moms}
\end{equation}%
where $\mathcal{L}_{p}^{-1}$ denotes the inverse Laplace transform with
respect to $p$, which is the Laplace conjugate to $N$. Carrying out the
inverse Laplace transform of (\ref{moms}) using Maple or Mathematica we have
analytically computed $28$ moments of $G(r,N)$ \cite{SS02}. From the
expansion of the exact formula (\ref{w8a}) for large $p$ at different orders
of truncations we have obtained the leading terms of $G(k,p)$ as
\begin{equation*}
\frac{1}{p}-\frac{1}{3p^{3}}k^{2}+\frac{1}{5p^{5}}k^{4}-\frac{1}{7p^{7}}%
k^{6}+...,
\end{equation*}%
which is nothing but the series of the Fourier-Laplace transform $%
G_{rod}(k,p)=k^{-1}\arctan (k/p)$ of the end-to-end distribution function of
a stiff rod $G_{rod}(r,N)=(4\pi N^{2})^{-1}\delta (r-N)$. Taking into
account the subleading terms in the expansion of $G(k,p)$
\begin{equation*}
\frac{1}{3p^{4}}k^{2}-\frac{2}{3p^{6}}k^{4}+\frac{1}{p^{8}}k^{6}-\frac{4}{%
3p^{10}}k^{8}+...
\end{equation*}%
results in the following expansion of the end-end-distribution function for
small $N$
\begin{equation}
G(r,N)=G_{rod}(r,N)-\frac{N}{6}\frac{d}{dN}G_{rod}(r,N)+...  \label{gr-N}
\end{equation}%
The latter can be considered as a singular expansion of the end-to-end
distribution function over its width. In principle, the expansion (\ref{gr-N}%
) can be extended to take into account the next terms. However, so far we
could not sum the expansion of the next to subleading terms in Eq.(\ref{w8a})%
\begin{equation*}
\frac{1}{p^{3}}(-\frac{1}{3}(\frac{k}{p})^{2}+\frac{29}{15}(\frac{k}{p})^{4}-%
\frac{86}{15}(\frac{k}{p})^{6}+\frac{38}{3}(\frac{k}{p})^{8}-\frac{71}{3}(%
\frac{k}{p})^{10}+\frac{119}{3}(\frac{k}{p})^{12}-...).
\end{equation*}

\subsection{The spectral representation of G(k,p)}

\label{spectral}

The spectral expansion of the matrix $A=DM$ enables one to derive the
representation of $\tilde{P}^{s}(k,p)$ through the eigenvalues and the
eigenvectors of the matrix $A$. The latter is useful in the treatment of
polymer adsorption, which is considered in Sect.\ref{ads-if}. The matrix $A$
is not symmetric, so that to find its spectral expansion one should consider
the eigenvalue problem for both $A$ and the transposed matrix $\overline{A}$%
. The eigenvalues of $A$ and $\overline{A}$ coincide, while their
(normalized) eigenvectors are different and are denoted by $u(n)$ and $%
\overline{u}(n)$, respectively. Thus, the spectral decomposition of $A$ reads%
\begin{equation}
A_{ij}=\sum\limits_{n}\lambda _{n}u_{i}(n)\overline{u}_{j}(n).  \label{w12}
\end{equation}%
Using (\ref{w12}) we obtain $\tilde{P}^{s}(k,p)$ as%
\begin{equation}
\left\langle l|\tilde{P}^{s}(k,p)|l^{\prime }\right\rangle =\sum\limits_{n}%
\frac{1}{1+ik\lambda _{n}}u_{l}(n)\overline{u}_{l^{\prime }}(n)D_{l^{\prime
},l^{\prime }}.  \label{w13}
\end{equation}%
The eigenvalues $\lambda _{n}$ of $A$ truncated by even order matrix build
pairs with the same absolute values and opposite sign in the pair, which we
denote by $\pm \overline{\lambda }_{k}$. For odd order $A$ one eigenvalue is
zero, and the remainder build pairs similar to those for even order
matrices. The largest eigenvalue of $A$ approaches the value $1/\sqrt{3p}$
for $p\rightarrow 0$ (the flexible limit). These properties of the
eigenvalues of $A$ enable one to write $G(k,p)$ as%
\begin{equation}
G(k,p)=\left\langle 0|\tilde{P}^{s}(k,p)|0\right\rangle =\frac{1}{2p}%
\sum\limits_{m}\frac{1}{1+k^{2}\lambda _{m}^{2}}l_{m},  \label{w14}
\end{equation}%
where we have introduced the notation $l_{m}=u_{0}(m)\overline{u}_{0}(m)$.
The summations in (\ref{w14}) has to be carried out over all eigenvalues. We
explicitly computed the eigenvalues and eigenvectors of $A$ with Maple or
Mathematica using truncations until $n=8$. The eigenvalues and the factors $%
l_{m}$ using truncation with four order matrices are obtained as%
\begin{equation*}
\overline{\lambda }_{1}=\sqrt{35}\frac{%
(p(18+27p+10p^{2}+p^{3})(85p+15p^{2}+105+\Delta ))^{1/2}}{%
630p+945p^{2}+350p^{3}+35p^{4}}
\end{equation*}%
\begin{equation*}
\overline{\lambda }_{3}=\sqrt{35}\frac{%
(p(18+27p+10p^{2}+p^{3})(85p+15p^{2}+105-\Delta ))^{1/2}}{%
630p+945p^{2}+350p^{3}+35p^{4}}
\end{equation*}%
with $\Delta =(7540p^{2}+1500p^{3}+15960p+120p^{4}+11025)^{1/2}$. \
\begin{equation*}
l_{1}=\frac{1}{12\Delta }(3\Delta -10p^{2}+60p+315),\ l_{3}=\frac{1}{%
12\Delta }(3\Delta +10p^{2}-60p-315)\
\end{equation*}%
The use of the spectral representation of $G(k,p)$ given by Eq.(\ref{w14})
enables one to carry out easily the inverse Fourier transform of $G(k,p)$,
and reduces the computation of $G(r,N)$ to performing the inverse Laplace
transform of $G(r,p)$. However, the computation of $G(r,N)$ using truncated
expressions results in $G(r,N)$ taking negative values for large $r$, and
demands a special consideration. Very recently the computation of $G(r,N)$
by using different methods was considered in \cite{hamprecht-kleinert}-\cite%
{winkler03}.

Notice the following difference in truncation of (\ref{w8}-\ref{w9}) with
odd and even order matrices. While $G(k,p)$ for even $n$ behaves for large $%
k $ as $1/k^{2}$, it behaves as $const$ using truncations with odd $n$. The
moment expansion as well the expansion for large $p$ of $G(k,p)$ behaves
correctly.

\subsection{\textbf{The Markovian property of the end-to-end distribution
function}}

\label{markov}

It is well-known that the end-to-end distribution function of an ideal
continuous flexible polymer (and of a polymer in an external potential)
possesses the Markovian property, which takes in terms of the Fourier
transform $P_{0}(k,N)=\exp (-k^{2}N/d)$ of the end-to-end distance, the
simple form, $P_{0}(k,N)=P_{0}(k,N-s)P_{0}(k,s)$. The Markovian property is
obviously not valid in this form for a semiflexible polymer. Nevertheless,
it can be generalized in an appropriate form to a semiflexible polymer, too.
Using the definition of the end-to-end distribution function (\ref{w1}) we
obtain in a straightforward way

\begin{eqnarray}
G(\mathbf{r}-\mathbf{r}_{0},N) &=&\int_{k}\left\langle \exp (i\mathbf{k}(%
\mathbf{r}-\mathbf{r}_{0})-i\mathbf{k}\int_{0}^{N}ds\mathbf{t}%
(s))\right\rangle  \notag \\
&=&\int d^{3}r^{\prime }\left\langle \int_{k2}\exp (i\mathbf{k}_{2}(\mathbf{r%
}-\mathbf{r}^{\prime })-i\mathbf{k}_{2}\int_{s^{\prime }}^{N-s^{\prime }}ds%
\mathbf{t}(s))\right.  \notag \\
&&\times \left. \int_{k1}\exp (i\mathbf{k}_{1}(\mathbf{r}^{\prime }-\mathbf{r%
}_{0})-i\mathbf{k}_{1}\int_{0}^{s^{\prime }}ds\mathbf{t}(s))\right\rangle ,
\label{w15}
\end{eqnarray}%
where $s^{\prime }$ fulfils the condition $0\leq s^{\prime }\leq N$.
Representing the average in (\ref{w15}) over the eigenstates of the rigid
rotator we arrive at
\begin{equation}
\left\langle 0|P^{s}(k,N)|0\right\rangle =\sum_{l}\left\langle
0|P^{s}(k,N-s)|l\right\rangle \left\langle l|P^{s}(k,s)|0\right\rangle ,
\label{w16}
\end{equation}%
where $P^{s}(k,N)$ is the inverse Laplace transform of the matrix $\tilde{P}%
^{s}(k,p)$. Eq.(\ref{w16}) is the generalization of the Markovian property
for a semiflexible polymer. Notice that the summation in (\ref{w16}) occurs
only over the quantum number of the square of the angular momentum $l$.
Similarly, one can show that the Markovian property applies to the
off-diagonal matrix elements $\left\langle l|P^{s}(k,N)|l^{\prime
}\right\rangle $, too. Eq.(\ref{w16}) can be immediately generalized by
partitioning the interval ($0$, $N$) in $\delta s=N/n$ intervals. The latter
is an important ingredient of our consideration of the behaviour of a
semiflexible polymer in half space (see Sec.\ref{dirichlet}).

Notice that the Markovian property of the quantity $\left\langle 0,0|P(%
\mathbf{k},N)|0,0\right\rangle $
\begin{equation}
\left\langle 0,0|P(\mathbf{k},N)|0,0\right\rangle =\sum_{l,m}\left\langle
0,0\right\vert P(\mathbf{k},N-s)\left\vert l,m\right\rangle \left\langle
l,m\right\vert P(\mathbf{k},s)\left\vert 0,0\right\rangle  \label{m-qm}
\end{equation}%
is the consequence of the group property of the time evolution operator of
the rigid rotator in an external field. Applying (\ref{mel1}) to both
transition amplitudes in (\ref{m-qm}), and using the addition theorem for
spherical functions gives again Eq.(\ref{w16}).

\subsection{Tangent correlations}

We now will consider two examples of computation of quantities containing
tangents. The correlation function of tangents $\left\langle \mathbf{t}(s)%
\mathbf{t}(s^{\prime })\right\rangle $ can be calculated using (\ref{w3}) as
\begin{equation*}
\left\langle \mathbf{t}(s)\mathbf{t}(s^{\prime })\right\rangle
=\sum\limits_{i=x,y,z}\sum\limits_{l,m}\left\langle 0,0\right\vert
t_{i}\left\vert l,m\right\rangle e^{-\frac{1}{2}l(l+1)\left\vert s-s^{\prime
}\right\vert }\left\langle l,m\right\vert t_{i}\left\vert 0,0\right\rangle
=e^{-\left\vert s-s^{\prime }\right\vert },
\end{equation*}%
where the arc length is measured in units of $l_{p}$.

The use of Eq.(\ref{mel1}) enables one to derive the following exact
expression for the Fourier transform of the end-to-end distribution function
with the fixed tangent $\mathbf{t}(0\mathbf{)}$
\begin{equation}
G(\mathbf{t,k};N)=\sum\limits_{l}\left\langle 0\mid P^{s}(k,N)\mid
l\right\rangle \sqrt{\frac{2l+1}{4\pi }}P_{l}(\mathbf{tn}).  \label{tng}
\end{equation}%
Eq.(\ref{tng}) enables one to compute the transversal moments (with respect
to the direction $\mathbf{t}(0\mathbf{)}$) of the free end of the polymer $%
\mathbf{r}(N)$. In the case if the tangent $\mathbf{t}(0)$ is parallel to
the $z$-axes, the scalar product $\mathbf{tn}$ in $P_{l}(\mathbf{tn})$ in (%
\ref{tng}) becomes zero. The calculation of the 2nd and the 4th transversal
moments using Eq.(\ref{tng}) in the case if $\mathbf{t}(0)\uparrow \uparrow
\mathbf{e}_{z}$ yield
\begin{equation*}
\left\langle r_{tr}^{2}\right\rangle =\frac{2}{9}\left(
6\,N-e^{-3\,N}+9e^{-N}-8\right) ,\,
\end{equation*}%
\begin{eqnarray*}
\left\langle r_{tr}^{4}\right\rangle &=&\frac{8}{9}(4\,N^{2}-\frac{248\,N}{15%
}+\frac{1834}{75}+\frac{3}{1225\,}e^{-10\,N}-\frac{1}{25\,}e^{-6\,N}+ \\
&&\frac{80}{147\,}e^{-3\,N}-\frac{624}{25\,}e^{-N}+\frac{4\,}{21\,}%
Ne^{-3\,N}-\frac{36\,}{5\,}Ne^{-N}).
\end{eqnarray*}%
The computation of higher moments\ is similar. For small $N$ the moments
behave as $\left\langle r_{tr}^{2}\right\rangle \simeq (2/3)N^{3}$, $%
\left\langle r_{tr}^{4}\right\rangle \simeq (8/9)N^{6}$. The $N^{3}$%
-dependence of $\left\langle r_{tr}^{2}\right\rangle $ on $N$ means that the
transversal fluctuations, which are controlled by the bending energy, are
small in comparison to the length of the polymer $N$. For large $N$ the
moments behave as $\left\langle r_{tr}^{2}\right\rangle \simeq $ $(4/3)N$, $%
\left\langle r_{tr}^{4}\right\rangle \simeq (32/9)N^{2}$, and obey the
relation between the moments of a Gaussian distribution in two dimensions, $%
\left\langle r_{tr}^{4}\right\rangle =2\left\langle r_{tr}^{2}\right\rangle
^{2}$. Note that his relation is also fulfilled for small $N$. The quotient $%
\left\langle r_{tr}^{4}\right\rangle /\left\langle r_{tr}^{2}\right\rangle
^{2}$ has the minimum equal to $1.36$ at $N=1.82$.

\section{The structure factor of a semiflexible polymer}

\label{s-fact}

\begin{figure}[tbph]
\includegraphics[clip,width=3in]{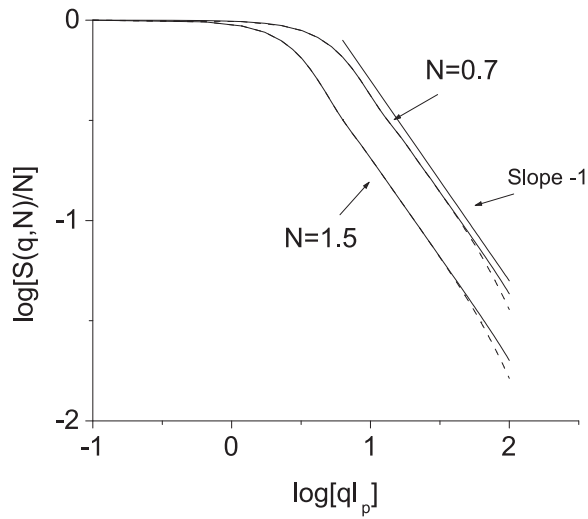}
\caption{The log-lot plot of the structure factor of a semiflexible polymer.
The dashed curves: truncations with number of exact moments $n=14$;
continuous curves: $n=18$. }
\label{sf-log}
\end{figure}

In studying the structure factor of a semiflexible polymer which is defined
by
\begin{equation}
S(q,N)=\frac{2}{N}\int_{0}^{N}ds_{2}\int_{0}^{s_{2}}ds_{1}\left\langle \exp
(i\mathbf{q}(\mathbf{r}(s_{2})-\mathbf{r}(s_{1})))\right\rangle .
\label{w17}
\end{equation}%
we first express $\mathbf{r}(s_{2})-\mathbf{r}(s_{1})$ in (\ref{w17})
through the tangent vectors, $\mathbf{r}(s_{2})-\mathbf{r}%
(s_{1})=\int_{s_{1}}^{s_{2}}ds\mathbf{t}(s)$. Representing the average in (%
\ref{w17}) through the eigenstates of quantum rigid rotator yields that the
structure factor of the semiflexible polymer $S(q,N)$ is the inverse Laplace
transform of $G(q,p)/p^{2}$ multiplied with the factor $2/N$ \cite{SS02}.
Fig.\ref{sf-log} shows the double logarithmic plot of the structure factor
of a semiflexible polymer as a function of the absolute value of the
scattering vector $q$ (measured in units of $l_{p}$) using truncations of
the exact matrix expression with finite order matrices.
The slope $-1$, which is characteristic for rigid rod behaviour, is also
shown as a guide for eyes. Note that the curves for $n=8$ and $n=10$ takes
exactly into account $14$ and $18$ moments of the end-to-end distribution
function, respectively.
\begin{figure}[tbph]
\includegraphics[clip,width=3in]{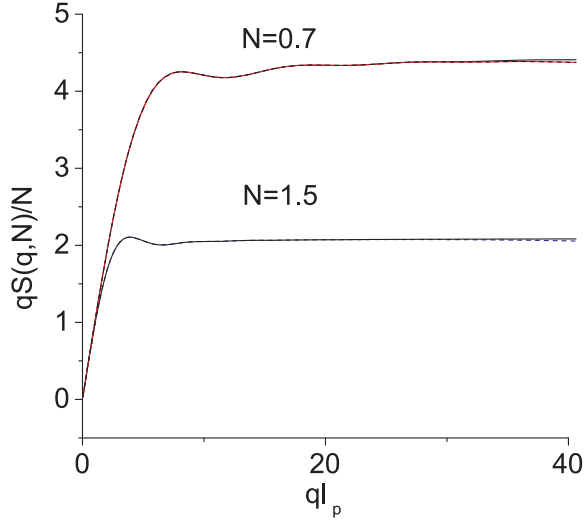}
\caption{The structure factor of a semiflexible polymer, $qS(q)$. Dashed
curves: $n=14$, continuous curves $n=18$. The structure factor of rigid rod
plotted for $N=0.7$ coincides practically with the dashed curve.}
\label{sf-q}
\end{figure}

The structure factor of a semiflexible polymer approaches for small $N$ that
of a stiff rod. The latter can be computed using $G(k,N)=\sin (qN)/qN$,
carrying out the inverse Laplace transform of $G(k,p)/p^{2}$, and
multiplying it with $2/N$. The result is%
\begin{equation*}
S(q,N)_{rod}=\frac{2}{Nq^{2}}(Nq\mathrm{Si}(qN)+\cos (qN)-1),
\end{equation*}%
where $\mathrm{Si}(x)=\int_{0}^{x}dt\sin (t)/t$. The subleading terms in the
expansion of $G(k,p)$ for large $p$ given by (\ref{gr-N}) result in the
following correction to the structure factor of the stiff rod for small $N$%
\begin{equation}
S_{1}(q,N)=\frac{2}{3q^{2}}-\frac{\sin \left( q\,N\right) }{q^{3}N}+\frac{%
\cos \left( q\,N\right) }{3q^{2}}.  \label{subl-sf}
\end{equation}%
The plots of $qS(q,N)$ and $qS(q,N)_{rod}$ are shown in Fig.\ref{sf-q}.

The computation of the structure factor is exact until the values of $q$
where the curves associated with different truncations begin to diverge.
Thus, the present method enables one in fact an exact computation of the
structure factor of the Kratky-Porod model without restriction on the
polymer length. The structure factor computed in \cite{yoshizaki-yamakawa80}
describes the low $q$ range corresponding to the Gaussian coil region. The
exact structure factor for the infinitely long chain given in \cite%
{descloizeaux73} is valid only for large $q$. The approximative approach
used in \cite{marques-fredrickson97} does not go beyond the exact
description of the exact 2nd moment of the end-to-end distribution function.
The expression of the structure factor obtained in \cite{kholodenko} gives
an interpolation between the Gaussian coil and stiff rod limits.

Besides the interest in its own as an experimentally accessible quantity,
the structure factor can be used within the random phase approximation \cite%
{degennes-book} to study the effects of rigidity on the phase behaviour of
polymer mixtures of different architectures \cite{binder}.

\section{Adsorption in a weak symmetric potential}

\label{ads-if}

We now will consider the adsorption of a semiflexible polymer in an external
delta-potential $U(z)=-u\delta (z)$, which corresponds to adsorption onto an
interface placed at $z=0$. The strength of the potential $u$ is measured in
units of $k_{B}T/l_{p}$. First we consider the partition function of the
polymer with ends fixed at $\mathbf{r}$ and $\mathbf{r}^{\prime }$ in a
general potential $U(\mathbf{r})$, which reads
\begin{equation}
Z(\mathbf{r},N;\mathbf{r}^{\prime })=\left\langle \delta (\mathbf{r}-\mathbf{%
r}(N))\delta (\mathbf{r}^{\prime }-\mathbf{r}(0))\exp (-\int_{0}^{N}dsU(%
\mathbf{r}(s)))\right\rangle .  \label{w22}
\end{equation}%
It can be shown in a straightforward way that the Taylor series of (\ref{w22}%
) in powers of the interaction potential can be written as

\begin{eqnarray}
Z(\mathbf{r},N;\mathbf{r}^{\prime }) &=&\int_{k}\left\langle \exp (i\mathbf{k%
}(\mathbf{r}-\mathbf{r}^{\prime })-i\mathbf{k}\int_{0}^{N}ds\mathbf{t}%
(s))\right\rangle +\sum\limits_{n=1}^{\infty }(-1)^{n}\int d^{3}r_{n}...\int
d^{3}r_{1}  \notag \\
&&\times \int_{q_{n}}...\int_{q_{1}}U(\mathbf{q}_{n})...U(\mathbf{q}%
_{1})\int_{k_{n+1}}...\int_{k_{1}}\exp (i\mathbf{q}_{n}\mathbf{r}_{n}+...+i%
\mathbf{q}_{1}\mathbf{r}_{1}  \notag \\
&&+i\mathbf{k}_{n+1}(\mathbf{r}-\mathbf{r}_{n})+i\mathbf{k}_{n}(\mathbf{r}%
_{n}-\mathbf{r}_{n-1})+...+i\mathbf{k}_{1}(\mathbf{r}_{1}-\mathbf{r}^{\prime
}))  \notag \\
&&\times \left\langle
\int_{0}^{N}ds_{n}\int_{0}^{s_{n}}ds_{n-1}...\int_{0}^{s_{2}}ds_{1}\exp (-i%
\mathbf{k}_{n+1}\int_{s_{n}}^{N}ds\mathbf{t}(s))\right.  \notag \\
&&\times \left. \exp (-i\mathbf{k}_{n}\int_{s_{n-1}}^{s_{n}}ds\mathbf{t}%
(s))...\exp (-i\mathbf{k}_{1}\int_{0}^{s_{1}}ds\mathbf{t}(s))\right\rangle .
\label{w23}
\end{eqnarray}%
Performing the integrations over $\mathbf{r}_{i}$ and $\mathbf{k}_{i}$
establishes the equivalence of Eq.(\ref{w23}) with the Taylor series of (\ref%
{w22}). Expressing the average in (\ref{w23}) through the eigenstates of the
rigid rotator, and carrying out the Laplace transform with respect to $N$ \
we obtain the bracket in the 2nd term of (\ref{w23}) as
\begin{eqnarray}
\left\langle ...\right\rangle &=&\left\langle 0,0|\tilde{P}(\mathbf{k}%
_{n+1},p)|l_{n},m_{n}\right\rangle \left\langle l_{n},m_{n}|\tilde{P}(%
\mathbf{k}_{n},p)|l_{n-1},m_{n-1}\right\rangle  \notag \\
&&...\left\langle l_{1},m_{1}|\tilde{P}(\mathbf{k}_{1},p)|0,0\right\rangle ,
\label{w24}
\end{eqnarray}%
where $p$ is Laplace conjugate to $N$, and $\tilde{P}(\mathbf{k}_{n},p)$ is
given by Eq.(\ref{w9}). The summations over $l_{i}$ and $m_{i}$ ($i=1$, $...$%
, $n$) are implied in (\ref{w24}).

The partition function $Z(z,p;z^{\prime })$ of the polymer in an adsorbing
potential $U(z)=-u\delta (z)$ can be obtained from Eq.(\ref{w23})
integrating it over $\mathbf{r}^{\shortparallel }$, so that $\mathbf{k}%
_{i}^{\shortparallel }$ in (\ref{w24}) become zero. The use of Eqs.(\ref{m})
permits to replace the matrices $\tilde{P}(\mathbf{k}_{n},p)$ in favor of
the square matrices $\tilde{P}^{s}(k_{n},p)$. As it is shown in Sect.\ref%
{g-cons} the factors $(\pm 1)^{l_{i}}$, where the sign minus corresponds to
the case when $k_{i}^{z}$ and $k_{i-1}^{z}$ have different sign, appear in
the intermediate states. It can be directly shown that the factors $%
(-1)^{l_{i}}$ can be taken into account by allowing $k_{i}$ in $\tilde{P}%
^{s}(k_{i},p)$ to take positive and negative values. Inserting the
delta-potential $U(z)=-u\delta (z)$ into Eq.(\ref{w23}) we obtain the
partition function of adsorbed polymer as
\begin{eqnarray}
Z(z,p;z^{\prime }) &=&\int_{k}\exp (ik(z-z^{\prime }))\left\langle 0|\tilde{P%
}^{s}(k,p)|0\right\rangle   \notag \\
&&+\sum\limits_{l_{1},l_{2}}\int_{k_{1}}\exp (ik_{1}z)\left\langle 0|\tilde{P%
}^{s}(k_{1},p)|l_{1}\right\rangle   \notag \\
&&\times u\left\langle l_{1}|(I-(u/2\pi )\int_{-\infty }^{\infty }dk\tilde{P}%
^{s}(k,p))^{-1}|l_{2}\right\rangle   \notag \\
&&\times \int_{k_{2}}\exp (-ik_{2}z^{\prime })\left\langle l_{2}|\tilde{P}%
^{s}(k_{2},p)|0\right\rangle ,  \label{w25}
\end{eqnarray}%
where $I$ is a diagonal matrix of infinite order. The poles of the partition
function are given by the zero points of the determinant
\begin{equation}
\det (I-(u/2\pi )\int_{-\infty }^{\infty }dk\tilde{P}^{s}(k,p))=0.
\label{w26}
\end{equation}%
Inserting $\tilde{P}%
^{s}(k,p)=(I+ikDM^{s})^{-1}D=(I+k^{2}(DM^{s})^{2})^{-1}(I-ikDM^{s})D$ into (%
\ref{w26}) we obtain the eigenvalue condition in the form $\det (I-(u/2\pi
)\int_{-\infty }^{\infty }dk(I+k^{2}(DM^{s})^{2})^{-1}D)=0$. We would like
to stress that Eqs.(\ref{w25}-\ref{w26}) are exact for adsorption in a
Dirac's delta-potential. The formula for the flexible polymer is obtained
from Eq.(\ref{w25}) by neglecting the off diagonal matrix elements and using
the expression $\tilde{P}^{s}(k,p)=1/(k^{2}/3+p)$ for the propagator.

It is interesting question if the energy eigenvalue condition (\ref{w26})
can be rewritten in terms of the boundary conditions imposed on "a wave
function" as it is the case for a flexible polymer. Such an interpretation
of Eqs.(\ref{w25}-\ref{w26}) would enable one to consider the adsorption in
potentials with finite widths and depths. An heuristic attempt to treat the
adsorption in this way was undertaken in \cite{stepanow-jcp01}. Notice that
according to the analogy of (\ref{w25}) with the corresponding equation for
a flexible polymers it is tempting to interpret $\int_{k_{1}}\exp
(ik_{1}z)\left\langle 0|\tilde{P}^{s}(k_{1},p)|l_{1}\right\rangle $ as the
"wave function" of the localized semiflexible polymer.

The integrations over $k$ in (\ref{w25}) can be carried out using the
spectral representation (\ref{w12}) according to%
\begin{equation*}
\int_{k}\exp (\pm ikz)\left\langle l_{1}|\tilde{P}^{s}(k,p)|l_{2}\right%
\rangle =\sum\limits_{m}f_{m}^{(\pm )}(z)u_{l_{1}}(m)\overline{u}%
_{l_{2}}(m)D_{l_{2},l_{2}}.
\end{equation*}%
with $f_{m}^{(\pm )}(z)=\int_{k}\exp (\pm ikz)(1-ik\lambda
_{m})/(1+k^{2}\lambda _{m}^{2})=\frac{\left\vert \lambda _{m}\right\vert \pm
\lambda _{m}}{2\lambda _{m}^{2}}\exp (-\left\vert z/\lambda _{m}\right\vert
) $. Note that the integral $\int_{-\infty }^{\infty }dzf_{m}^{(\pm )}(z)$
is equal to $1$. The inverse Laplace transform of (\ref{w25}), which takes
into account for large $N$ the main contributions associated with poles
defined by Eq.(\ref{w26}), yields
\begin{eqnarray}
Z(z,N;z^{\prime })_{bound} &\sim
&\sum\limits_{p_{0}}e^{p_{0}N}\sum\limits_{l_{1},l_{2}}\sum%
\limits_{m}f_{m}^{(+)}(z)u_{0}(m)\overline{u}_{l_{1}}(m)D_{l_{1},l_{1}}
\notag \\
&&\times \frac{A_{l_{1},l_{2}}(p_{0})}{\delta (p_{0})}\sum%
\limits_{n}f_{n}^{(-)}(z^{\prime })u_{l_{2}}(n)\overline{u}_{0}(n)D_{0,0},
\label{w25a}
\end{eqnarray}%
where%
\begin{equation*}
\delta (p_{0})=\lim_{p\rightarrow p_{0}}\det (I-(u/2\pi )\int_{-\infty
}^{\infty }dk\tilde{P}^{s}(k,p))/(p-p_{0}),
\end{equation*}%
and $A_{l_{1},l_{2}}(p_{0})$ is the adjoint of the matrix element $%
\left\langle l_{1}|(I-(u/2\pi )\int_{-\infty }^{\infty }dk\tilde{P}%
^{s}(k,p))^{-1}|l_{2}\right\rangle $ taken at $p_{0}$, which is the zero of (%
\ref{w26}). Eq.(\ref{w25a}) is the spectral expansion of the partition
function over the localized states, and shows that the zeros $p_{0}$ of Eq.(%
\ref{w26}) yield the energy spectrum of the localized semiflexible polymer.
The distribution function for monomers $n(z)$ is calculated using the
expression
\begin{equation}
n(z)=\frac{1}{\int dz\int dz^{\prime }Z(z,N;z^{\prime })}\int_{0}^{N}ds\int
dz^{\prime }\int dz^{\prime \prime }Z(z^{\prime
},N-s;z)_{0,l}Z(z,s;z^{\prime \prime })_{l,0},  \label{n-za}
\end{equation}%
which generalizes the corresponding formula of flexible polymer \cite%
{degennes69}. The quantity $Z(z^{\prime },N;z)_{0,l}$ is obtained from the
expression on the right-hand side of (\ref{w25a}) by replacing the index $0$
in $\overline{u}_{0}(n)$ and in $D_{0,0}$ by $l$. \ Further, in computing $%
n(z)$ we will take into account only the ground state (ground state
dominance). Inserting (\ref{w25a}) into (\ref{n-za}) results in the
following expression for the monomer density of adsorbed polymer%
\begin{equation}
n(z)=\sum\limits_{n,m}\sum\limits_{l_{2},l,l_{1}}\frac{1}{\delta (p_{0})}%
A_{0,l_{2}}(p_{0})f_{n}^{(-)}(z)u_{l_{2}}(n)\overline{u}%
_{l}(n)D_{l,l}f_{m}^{(+)}(z)u_{l}(m)\overline{u}%
_{l_{1}}(m)D_{l_{1},l_{1}}A_{l_{1},0}(p_{0}).  \label{n-z}
\end{equation}%
The distribution function of one polymer end of adsorbed polymer, which is
defined by%
\begin{equation*}
f(z)=\int dz^{\prime }Z(z,N;z^{\prime })/\int dz\int dz^{\prime
}Z(z,N;z^{\prime }),
\end{equation*}%
results after inserting (\ref{w25a}) and using the approximation of ground
state dominance in the following expression%
\begin{equation}
f(z)=\sum\limits_{l,n}f_{n}^{(+)}(z)u_{0}(n)\overline{u}%
_{l}(n)D_{l,l}A_{l,0}(p_{0})/A_{0,0}(p_{0})/D_{0,0}.  \label{f-z}
\end{equation}

We have mentioned in Sect.\ref{spectral} that $G(k,p)$ behaves differently
for large $k$ in using truncations with odd or even order matrices. This is
now important because Eq.(\ref{w26}) contains integrals over $k$ of $\tilde{P%
}^{s}(k_{1},p)$. Notice that the matrix elements in (\ref{w26}) which are
odd in $k$ vanish because the integration is carried out in symmetric
limits. While $G(k,p)$, which corresponds to an odd truncation, does not
have well-defined inverse Fourier transform with respect to $k$ due to the
divergence of the integral for large $k$, we will use truncations with even
order matrices. At present we do not have a more reasonable explanation for
this behaviour of $G(k,p)$ for odd and even $n$ for large $k$.

Thus, we will use in Eq.(\ref{w26}) truncations with even order matrices,
and will study the energy eigenvalue condition for $n=4$ and $n=6$, where $6$
and $10$ moments of the end-to-end distribution function are exact,
respectively. The use of the spectral decomposition of the matrix $DM^{s}$
according to (\ref{w12}) permits to perform easily the integration over $k$.
As a result we have obtained from (\ref{w26}) the following relations
between the adsorption strength $u$ and the localization energy $E$ in the
vicinity of the localization transition
\begin{equation}
u=\frac{2}{\sqrt{3}}\sqrt{-E}-\frac{8}{9}\sqrt{\frac{14}{5}}(-E)-\sqrt{3}%
\frac{2939}{405}(-E)^{3/2}+O((-E)^{2}),\ \ \ \ \ n=4  \label{w27}
\end{equation}%
\begin{equation}
u=\frac{2}{\sqrt{3}}\sqrt{-E}-2.264(-E)-2.0854(-E)^{3/2}+O((-E)^{2}),\ \ \ \
\ \ n=6  \label{w27a}
\end{equation}%
The 1st term on the right side of Eqs.(\ref{w27}-\ref{w27a}) is the result
for a flexible polymer written in units under consideration, where the
dimensionless adsorption strength and the energy are measured in units of $%
k_{B}T/l_{p}$ and $l_{p}$, respectively. The relation $l=\sqrt{2}l_{p}$
allows to replace the persistence length in favor of the statistical segment
length $l$. Eqs.(\ref{w27}-\ref{w27a}) show that the applicability of
adsorption theory of flexible polymer is in fact restricted to the
localization transition $u\rightarrow 0$. For all finite $u$ there are
corrections due to the polymer stiffness.

The normalized monomer density at the localization energy $-E\equiv
p_{0}=0.8 $, which is computed from Eq.(\ref{n-za}) using truncation with
matrices of order $4$, reads
\begin{equation}
n(z)=0.42e^{-3.84z}+2.35e^{-12.78z}+4.52e^{-21.73z}.  \label{n-z4}
\end{equation}%
In contrast, the monomer density for a flexible polymer for the same
localization energy is
\begin{equation*}
n_{fl}(z)=1.55e^{-3.10z}.
\end{equation*}%
Eq.(\ref{n-z4}) shows that the decay of the monomer density of adsorbed
semiflexible polymer is not exponential. This is also seen in Fig.\ref%
{fig-n-z} displaying the logarithmic plot of the monomer density for two
different localization energies. The monomer density for flexible polymer is
also shown for comparison.
\begin{figure}[tbph]
\includegraphics[clip,width=3in]{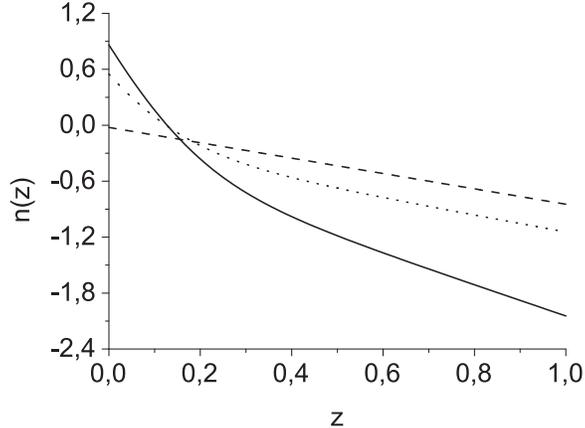}
\caption{The normalized monomer density of adsorbed polymer. Solid line:
adsorption energy $p=0.8$, adsorption strength $u=0.375$; dots: $p=0.3$, $%
u=0.305$; dashes: flexible polymer at $p=0.3$, $u=2p/\protect\sqrt{3}=0.346$%
. }
\label{fig-n-z}
\end{figure}
As it follows from Eq.(\ref{n-z}) and from the expression of the factor $%
f_{m}^{(\pm )}(z)$ the decay of $n(z)$ away from the interface is controlled
by the length $\left\vert \lambda _{1}\right\vert /2$, where $\lambda _{1}$
is the largest eigenvalue of the matrix $DM^{s}$. At $z$ smaller than $%
z_{c}\approx 0.22$, which is determined by the condition that the first term
in (\ref{n-z4}) have the same value as the rest, the decay of $n(z)$ is
sharper, and is determined by the length $\left\vert \lambda _{1}\lambda
_{2}\right\vert /(\left\vert \lambda _{1}\right\vert +\left\vert \lambda
_{2}\right\vert )$, where $\lambda _{2}$ is the next largest eigenvalue of $%
DM^{s}$. The behaviour for $z<z_{c}$ is likely to be ascribed to formation
of the liquid-crystalline ordering of the polymer caused by alignment of the
pieces of adsorbed polymer along the interface. The length $z_{c}$
separating the faster and the slower decays, can be interpreted as the
correlation length of the liquid-crystalline ordering induced by the
interface. The analysis of the behaviour of $n(z)$ for $n=4$ shows that in
approaching the localization transition $p_{0}\rightarrow 0$ the prefactors
in front of the 2nd and 3rd exponents in Eq.(\ref{n-z4}) tend to zero faster
than that in front of the first term, which tends to the value $\sqrt{3p_{0}}
$, which is the result of the flexible polymer. For truncations with $%
n\times n$ matrices the monomer density $n(z)$ is according to Eq.(\ref{n-z}%
) a superposition of $n(n+2)/8$ exponents. We expect that similar to the
case $n=4$ the monomer density tends to that of the flexible polymer in
approaching the localization transition, $p_{0}\rightarrow 0$.
\begin{figure}[tbph]
\includegraphics[clip,width=3in]{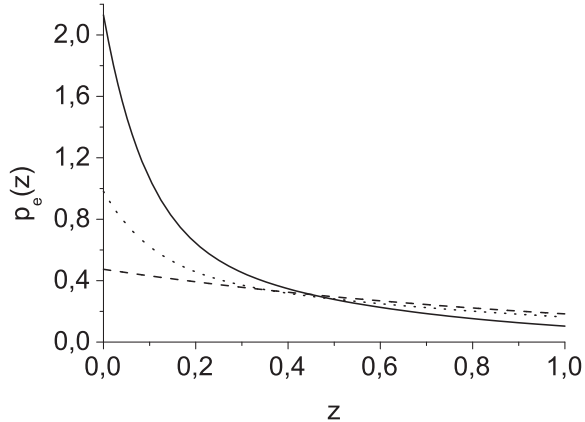}
\caption{The normalized distribution of one polymer end of adsorbed polymer.
Solid line: adsorption energy $p=0.8$; dots: $p=0.3$; dashes: flexible
polymer at $p=0.3$.}
\label{fig-f-end-z}
\end{figure}

It is surprising that the leading correction in Eqs.(\ref{w27}-\ref{w27a})
to adsorption of flexible polymer depends on the order of truncations. We
expect that at given adsorption strength there are pieces of the polymer
with characteristic size depending on $u$, which are completely adsorbed and
are aligned along the interface. The order of truncation should be such that
the statistics of these pieces is described accurately. With increasing $u$
the size of these pieces increases too, which demands more accurate
description by using the higher order matrices. However, the extreme
narrowness of the delta-potential demands high accuracy on quite small
scales, which is apparently the reason why the leading correction is
different for $n=4$ and $n=6$. Fig.\ref{fig-f-end-z} shows the distribution
function of one polymer end of adsorbed polymer computed using Eq.(\ref{f-z}%
). Similar to Fig.\ref{fig-n-z} we see here different dependence on $z$ for
small and large $z$.

Note that truncation of Eq.(\ref{w26}) with $2\times 2$ matrices, where only
the 2nd moment of the end-to-end distribution function is taken exactly into
account, yields the leading correction to the flexible limit as $(-E)^{3/2}$
instead of $(-E)$ in accordance with Eq.(\ref{w27}). Thus, truncation with $%
n=2$ does not correctly describe the leading correction to the flexible
polymer.

Eq.(\ref{w27}) shows that the correction term is negative, i.e. the
localization energy for semiflexible polymer at the same strength of the
attraction potential is higher than that for flexible one. This conclusion
is in agreement with preceding studies \cite{birshtein79}- \cite%
{kuznetsov/sung}. Adsorption of a semiflexible polymer considered in \cite%
{maggs89}-\cite{benetatos-frey}, was carried out in the framework of the
model, which is semiflexible on all scales. This model cannot describe the
regime of weak adsorption considered in the present work.\

\section{A semiflexible polymer in half space}

\label{dirichlet}

The Fourier transform of the end-to-end distribution function (\ref{w8a})
can be written in terms of the cumulants $\int_{0}^{N}ds\mu _{2n}(s)$ of the
moments as%
\begin{equation*}
G(\mathbf{k},N)=\exp (-\sum_{n=1}^{\infty }\int_{0}^{N}ds\mu _{2n}(s)(%
\mathbf{k}^{2})^{n}).
\end{equation*}%
It is easy to see that the latter results in the following differential
equation for the end-to-end distribution function%
\begin{equation}
\frac{\partial \ G(\mathbf{r-r}_{0},N)}{\partial N}-\sum_{n=1}^{\infty
}(-1)^{n+1}\mu _{2n}(N)\Delta _{\mathbf{r}}^{n}G(\mathbf{r-r}_{0},N)=0,
\label{xx}
\end{equation}%
where $\Delta =\nabla ^{2}$ is the Laplace operator. Eq.(\ref{xx}), which is
exact, enables one to derive the partition function of a semiflexible
polymer in half space $z\geq 0$ with Dirichlet condition at the boundary.
Taking into account that $G(r=\sqrt{(\mathbf{r}^{\shortparallel }-\mathbf{r}%
_{0}^{\shortparallel })^{2}+(z+z_{0})^{2}},N)$ also obeys Eq.(\ref{xx}), the
partition function of the semiflexible polymer in half space with Dirichlet
boundary condition at $z=0$ can be written as%
\begin{equation}
Z(\mathbf{r},\mathbf{r}_{0},N)=G(\sqrt{(\mathbf{r}^{\shortparallel }-\mathbf{%
r}_{0}^{\shortparallel })^{2}+(z-z_{0})^{2}},N)-G(\sqrt{(\mathbf{r}%
^{\shortparallel }-\mathbf{r}_{0}^{\shortparallel })^{2}+(z+z_{0})^{2}},N).
\label{w32a}
\end{equation}%
However, it is not clear from the above derivation if the trajectories of
the polymer associated with $Z(\mathbf{r},\mathbf{r}_{0},N)$ obey the
Dirichlet boundary condition at intermediate points, $0<s<N$. The following
derivation of (\ref{w32a}) based on the use of the Markovian property of the
end-to-end distribution function enables one to prove this. Partitioning the
interval $(0$, $N)$ in $n+1$ equal intervals $\delta s=N/(n+1)$ we obtain
\begin{eqnarray}
G(\mathbf{r}-\mathbf{r}_{0},N) &=&\int d^{3}r_{n}...\int
d^{3}r_{1}\int_{k_{n+1}}...\int_{k_{1}}\times   \notag \\
&&\exp (i\mathbf{k}_{n+1}(\mathbf{r}-\mathbf{r}_{n})+i\mathbf{k}_{n}(\mathbf{%
r}_{n}-\mathbf{r}_{n-1})+...+i\mathbf{k}_{1}(\mathbf{r}_{1}-\mathbf{r}_{0}))
\notag \\
&&\left\langle
0|P^{s}(k_{n+1},N-s_{n})P^{s}(k_{n},s_{n}-s_{n-1})...P^{s}(k_{1},s_{1})|0%
\right\rangle .  \label{w28}
\end{eqnarray}%
Notice that $k_{i}$ in the argument of $P^{s}(k_{i},s_{i}-s_{i-1})$ is the
absolute value of $\mathbf{k}_{i}$. The dependencies on positions $\mathbf{r}%
_{m}$ in Eq.(\ref{w28}) are the same as in the corresponding equation for a
flexible polymer, which is obtained from (\ref{w28}) replacing the factors $%
\tilde{P}^{s}(k,p)$ by $P_{fl}(k,N)=\exp (-k^{2}N/3)$, and taking into
account in the intermediate states only the term with $l=0$. In order to
obtain from (\ref{w28}) the partition function of the polymer in half space
we will proceed in the same manner as for flexible polymers, and replace the
factors $\exp (i\mathbf{k}_{m}(\mathbf{r}_{m}-\mathbf{r}_{m-1}))$ in (\ref%
{w28}) for $m=1,...,n$ according to
\begin{eqnarray}
\exp (i\mathbf{k}_{m}(\mathbf{r}_{m}-\mathbf{r}_{m-1})) &\rightarrow &\frac{1%
}{2}\exp (i\mathbf{k}_{m}^{\shortparallel }(\mathbf{r}_{m}^{\shortparallel }-%
\mathbf{r}_{m-1}^{\shortparallel }))  \notag \\
&&\times (\exp (i(k_{m}^{z}(z_{m}-z_{m-1}))-\exp
(i(k_{m}^{z}(z_{m}+z_{m-1})))).  \label{w29}
\end{eqnarray}%
For $m=n+1$ we use the replacement (\ref{w29}) without the factor $1/2$.
Inserting the latter into Eq.(\ref{w28}) gives the partition function of the
semiflexible polymer with fixed ends in the presence of a surface at $z=0$ as%
\begin{eqnarray}
Z(\mathbf{r},\mathbf{r}_{0},N) &=&\int d^{3}r_{n}...\int
d^{3}r_{1}\int_{k_{n+1}}...\int_{k_{1}}\exp (i\mathbf{k}_{n+1}^{%
\shortparallel }(\mathbf{r}^{\shortparallel }-\mathbf{r}_{n}^{\shortparallel
})+i\mathbf{k}_{n}^{\shortparallel }(\mathbf{r}_{n}^{\shortparallel }-%
\mathbf{r}_{n-1}^{\shortparallel })+...  \notag \\
&&+i\mathbf{k}_{1}^{\shortparallel }(\mathbf{r}_{1}^{\shortparallel }-%
\mathbf{r}_{0}^{\shortparallel }))2^{-n}(\exp (ik_{n+1}^{z}(z-z_{n}))-\exp
(ik_{n+1}^{z}(z+z_{n})))...\times   \notag \\
&&(\exp (ik_{1}^{z}(z_{1}-z_{0}))-\exp (ik_{1}^{z}(z_{1}+z_{0})))\times
\notag \\
&&\left\langle
0|P^{s}(k_{n+1},N-s_{n})P^{s}(k_{n},s_{n}-s_{n-1})...P^{s}(k_{1},s_{1})|0%
\right\rangle .  \label{w30}
\end{eqnarray}%
$Z(\mathbf{r},\mathbf{r}_{0},N)$ becomes zero, if $z$, $z_{0}$ or any
intermediate coordinate $z_{m}$ ($m=1,...,n$) are zero. Thus, the expression
(\ref{w30}) for $Z(\mathbf{r},\mathbf{r}_{0},N)$ fulfils in the limit $%
n\rightarrow \infty $ the Dirichlet boundary condition $z(s)=0$ ($0\leq
s\leq N$), and consequently Eq.(\ref{w30}) gives the partition function of
the semiflexible polymer with both ends fixed in the presence of a wall at $%
z=0$.

We now will show that in the limit $\delta s\rightarrow 0$ Eq.(\ref{w30})
passes over to Eq.(\ref{w32a}). To arrive at this result we carry out
successively integrations over $\mathbf{r}_{m}$ in (\ref{w30}). The
integration over $\mathbf{r}_{m}^{\shortparallel }$ gives $(2\pi
)^{d-1}\delta (\mathbf{k}_{m}^{\shortparallel }-\mathbf{k}%
_{m+1}^{\shortparallel })$ while the integration over $z_{m}$ yields
\begin{eqnarray}
&&2\pi (\exp (ik_{m+1}^{z}z_{m+1}-ik_{m}z_{m-1})\delta
(k_{m+1}^{z}-k_{m}^{z})-\exp (ik_{m+1}^{z}z_{m+1}+ik_{m}z_{m-1})\delta
(k_{m+1}^{z}-k_{m}^{z})-  \notag \\
&&\exp (ik_{m+1}^{z}z_{m+1}-ik_{m}z_{m-1})\delta
(k_{m+1}^{z}+k_{m}^{z})+\exp (ik_{m+1}^{z}z_{m+1}+ik_{m}z_{m-1})\delta
(k_{m+1}^{z}+k_{m}^{z}))  \label{w31}
\end{eqnarray}%
In obtaining (\ref{w31}) we have taken into account that $%
P^{s}(k_{m},s_{m}-s_{m-1})$ depends on the absolute value of $\mathbf{k}_{m}$%
, and consequently can be written as common factor. Using the Markovian
property of the distribution function (\ref{w16}) at the interval ($%
s_{m+1},s_{m-1}$), we obtain again the expression (\ref{w30}) with the
difference that the point $s_{m}$ is now missed, and the prefactor in front
of (\ref{w30}) will be $2^{-n+1}$. Repeating this procedure $n$ times we
finally obtain
\begin{eqnarray}
Z(\mathbf{r},\mathbf{r}_{0},N) &=&\int_{k}\exp (i\mathbf{k}^{\shortparallel
}(\mathbf{r}^{\shortparallel }-\mathbf{r}_{0}^{\shortparallel }))\left( \exp
(ik^{z}(z-z_{0}))-\exp (ik^{z}(z+z_{0}))\right) \left\langle
0|P(k,N)|0\right\rangle  \notag \\
&=&G(\sqrt{(\mathbf{r}^{\shortparallel }-\mathbf{r}_{0}^{\shortparallel
})^{2}+(z-z_{0})^{2}},N)-G(\sqrt{(\mathbf{r}^{\shortparallel }-\mathbf{r}%
_{0}^{\shortparallel })^{2}+(z+z_{0})^{2}},N).  \label{w32}
\end{eqnarray}%
The above derivation guarantees that the trajectories of the polymer $%
\mathbf{r}(s)$ contacting the surface at an arbitrary point $s\in (0,N)$ do
not contribute to $Z(\mathbf{r},\mathbf{r}_{0},N)$.

The derivation of (\ref{w30}-\ref{w32}) is based on the Markovian property
of the end-to-end distribution function given by Eq.(\ref{m-qm}), and the
dependence of $P^{s}(k,N)$ on the absolute value of the wave vector $\mathbf{%
k}$. The statistical weights of configurations, which are determined by the
expression in the brackets in Eq.(\ref{w30}) and are different for flexible
and semiflexible polymers, do not play a direct role in the above
derivation. The circumstance that the subtracted terms, which result in the
2nd term of Eq.(\ref{w32}), obey the infinite order differential equation of
the free polymer (\ref{xx}), guarantess that the polymer configurations
contributing to $Z(\mathbf{r},\mathbf{r}_{0},N)$ are those of the free
polymer. Note that Eqs.(\ref{w32a},\ref{w32}) are also valid for stiff rod, $%
\left\vert \mathbf{r}-\mathbf{r}_{0}\right\vert =N$, where the second term,
which is not zero only if $z$ or $z_{0}$ are zero, selects the
conformations, which do not have contact with the boundary.

\subsection{Behaviour of a semiflexible polymer in the vicinity of a wall}

\label{wall}

Eq.(\ref{w32}) enables one to study the behaviour of a semiflexible polymer
in the vicinity of a wall. For this goal we will use the approximative but
simple analytic form of the end-to-end distribution function derived in \cite%
{bhattacharjee97}%
\begin{equation}
G_{0}(r,L)=\frac{1}{(1-r^{2})^{9/2}}\exp (-\frac{9L}{8l_{p}(1-r^{2})}),
\label{w33}
\end{equation}%
where $r$ in Eq.(\ref{w33}) and throughout this subsection is measured in
units of the contour length $L$, so that as a consequence that the polymer
chain is locally inextensible, the distance $r$ fulfils the inequality $%
r\leq 1$. The computations can also be carried out with end-to-end
distribution functions derived recently in \cite{hamprecht-kleinert},\cite%
{winkler03}.

\begin{figure}[tbph]
\includegraphics[clip,width=3in]{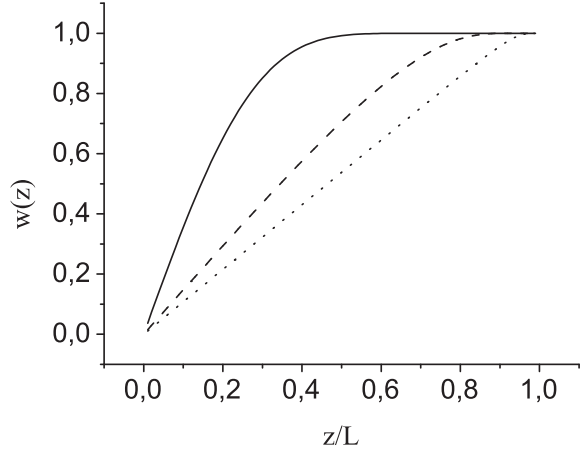}
\caption{The distribution function of one polymer end as a function of the
distance to the wall for different persistence lengths. Solid line: $%
L/l_{p}=10$; dashes: $L/l_{p}=1.5$; dots: $L/l_{p}=0.3$.}
\label{fig4}
\end{figure}

To compute the distribution function $G(\mathbf{r},\mathbf{r}_{0},N)$ with
one end of the polymer fixed at distance $z$ to the wall, we integrate Eq.(%
\ref{w32}) over $\mathbf{r}^{\shortparallel }-\mathbf{r}_{0}^{\shortparallel
}$ and $z_{0}$, and obtain%
\begin{equation}
w(z)=\pi \int_{1-z^{2}}^{1}\frac{d\beta }{\sqrt{1-\beta }}\int_{0}^{\beta
}dxg_{0}(x),  \label{w34}
\end{equation}%
where $g_{0}(x)$ is the normalized $G_{0}(r,L)$ and $x=1-r^{2}$. The
integration over $x$ can be performed analytically, while the integration
over $\beta $ only numerically. Fig.\ref{fig4} shows $w(z)$ for different
ratios $L/l_{p}$. For $z=1$ the polymer just begin to contact the wall, so
that $w(z=1)=1$. For large $L/l_{p}$ the semiflexible polymer is a coil with
the size\ which is proportional to $\sqrt{L}$, so that the effect of the
wall on $w(z)$ appears at smaller $z$. For small $L/l_{p}$ the polymer
behaves as stiff rod, and $w(z)$ decreases linearly at small $z$. Fig.\ref%
{fig4} shows that at given distance of the polymer end to the wall $z$, and
for given contour length $L$, the distribution function $w(z)$ is larger for
polymers which are more flexible.

\begin{figure}[tbph]
\includegraphics[clip,width=3in]{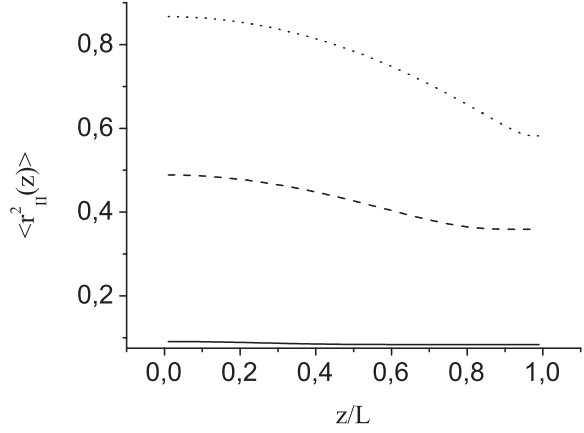}
\caption{The in-plane mean-square end-to-end distance of the polymer as a
function of the distance of one polymer end to the wall for different
persistence lengths. Solid line: $L/l_{p}=10$; dashes: $L/l_{p}=1.5$; dots: $%
L/l_{p}=0.3$.}
\label{fig5}
\end{figure}

To compute the projection of the mean-square end-to-end polymer distance
parallel to the wall under the condition that one end of the polymer is
fixed at distance $z$ to the wall we average $(\mathbf{r}^{\shortparallel }-%
\mathbf{r}_{0}^{\shortparallel })^{2}$ using Eqs.(\ref{w32}-\ref{w33}). The
normalization of the result with $w(z)$ gives%
\begin{equation}
\left\langle (\mathbf{r}^{\shortparallel }-\mathbf{r}_{0}^{\shortparallel
})^{2}\right\rangle _{z}=\pi \int_{1-z^{2}}^{1}\frac{d\beta }{\sqrt{1-\beta }%
}\int_{0}^{\beta }dx(\beta -x)g_{0}(x)/w(z).  \label{w35}
\end{equation}%
Fig.\ref{fig5} shows the mean-square end-to-end distance parallel to the
wall as a function of $z$ for different values of $L/l_{p}$ computed using
Eq.(\ref{w35}). While for Gaussian polymer the dependencies on longitudinal
and transverse distances separate, $\left\langle (\mathbf{r}^{\shortparallel
}-\mathbf{r}_{0}^{\shortparallel })^{2}\right\rangle _{z}$ does not depend
on $z$. In rod limit, $L/l_{p}\ll 1$, $\left\langle (\mathbf{r}%
^{\shortparallel }-\mathbf{r}_{0}^{\shortparallel })^{2}\right\rangle _{z}$
is a linear function having the values $1$ and $2/3$ at $z=0$ and $z=1$,
respectively. The approximate distribution function (\ref{w33}) reproduces
qualitatively the rod behaviour.

\begin{figure}[tbph]
\includegraphics[clip,width=3in]{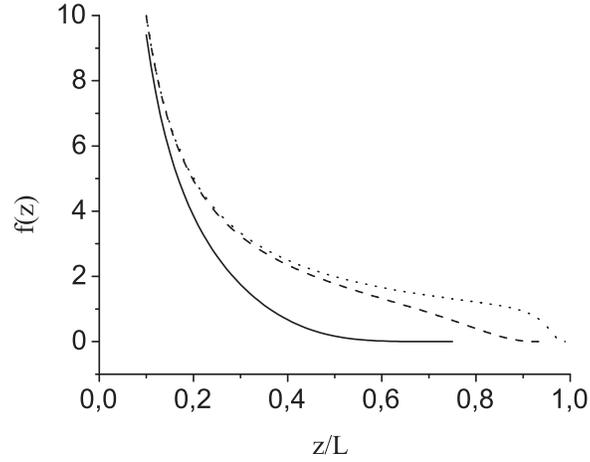}
\caption{The force acting on the wall as a function of the distance to the
wall for different persistence lengths. Solid line: $L/l_{p}=10$; dashes: $%
L/l_{p}=1.5$; dots: $L/l_{p}=0.3$.}
\label{fig6}
\end{figure}

Fig.\ref{fig6} shows the force $f(z)=\partial \ln w(z)/\partial z$ acting on
the polymer as a function of the distance of one polymer end to the wall. As
in the case of the distribution function $w(z)$ the force begins to deviate
from zero for more flexible polymer at smaller $z$. For very stiff polymer
there are two regimes in the behaviour of the force. Just below $z\leq 1$
the force increase is sharp, and becomes weaker with further decrease of $z$%
. We attribute this initial sharp increase of the force to the bending of
the rod.
It is intuitively clear that if the polymer just begin to contact the
surface, $z\leq 1$, the bending gives the major contribution to the force. \
With further decrease of $z$ the configurations with large bending are less
probable, and the increase of the force is of entropic origin.

\section{Adsorption of a semiflexible polymer onto a surface}

\label{ads-surf}

In considering the adsorption of a semiflexible polymer in a weak surface
potential defined by $U(z))=-u\delta (z-z_{0})$ for $z>0$ and $U(z))=\infty $
for $z\leq 0$, where the attractive delta-potential is placed at the
distance $z_{0}$ to the wall, we chose as a reference state the distribution
function in half space (\ref{w32}), and expand the partition function in
Taylor series in powers of the attractive part of the potential. As a result
we obtain the Laplace transform of the partition function of the polymer $%
Z(z,p;z^{\prime })$ as
\begin{eqnarray}
Z(z,p;z^{\prime }) &=&\int_{k}\left( \exp (ik(z-z^{\prime }))-\exp
(ik(z+z^{\prime }))\right) \left\langle 0|\tilde{P}^{s}(k,p)|0\right\rangle
\notag \\
&&+\int_{k_{1}}\left( \exp (ik_{1}(z-z_{0}))-\exp (ik_{1}(z+z_{0}))\right)
\notag \\
&&\times \int_{k_{2}}\left( \exp (ik_{2}(z_{0}-z^{\prime }))-\exp
(ik_{2}(z_{0}+z^{\prime }))\right) u\left\langle 0|\tilde{P}%
^{s}(k_{1},p)\right.  \notag \\
&&\times \left. (I-u\int_{k}(1-\exp (2ikz_{0})\tilde{P}^{s}(k,p))^{-1}\tilde{%
P}^{s}(k_{2},p)|0\right\rangle .  \label{w36}
\end{eqnarray}%
The procedure of Sec.\ref{dirichlet} which we used to derive Eq.(\ref{w32})
starting with (\ref{w30}) guarantees that in expressions like $%
\int_{k}\left( \exp (ik(z-z_{0}))-\exp (ik(z+z_{0}))\right) \tilde{P}%
^{s}(k,N)$ entering (\ref{w36}), where $k$ in $\tilde{P}^{s}(k,N)$ takes
both positive and negative values, the Dirichlet boundary condition at $z=0$
is correctly taken into account along the contour length of the polymer.

The poles of the partition function, which are the zero points of the
determinant%
\begin{equation}
\det (I-u\int_{k}(1-\exp (2ikz_{0})\tilde{P}^{s}(k,p))=0  \label{w37}
\end{equation}%
gives the localization energy of adsorbed polymer. The truncation of $\tilde{%
P}^{s}(k,p)$ by a finite order matrix permits to study the adsorption of a
semiflexible polymer in a week surface potential. Truncation with matrix of
order $n$ gives the eigenvalue condition from (\ref{w37}) as a polynomial of
nth degree in powers of $u$. Due to the same reasons as for adsorption in a
symmetric potential we will evaluate (\ref{w37}) using truncations with even
size matrices. We will study here only the effect of polymer stiffness on
the threshold value of the strength of \ the adsorbing potential $u$, i.e.
Eq.(\ref{w37}) for $p=0$. For a flexible polymer the threshold value of the
potential strength $u_{c}$, such that for $u<u_{c}$ the polymer is
delocalized, obeys the condition $1-3u_{c}z_{0}=0$. \ The computation using
truncation with $4\times 4$ matrices at the value $z_{0}=1/3$ gives the
critical value of the localization strength $u_{c}=0.3285$, which is smaller
than $u_{c}=1$ for flexible polymer, and shows that the semiflexible polymer
adsorbs easier than the flexible one.

\section{Semiflexible polymers with monomer-monomer interactions}

\label{self-inter}

The formalism developed in this work enables one to study semiflexible
polymers with monomer-monomer interactions which are described by the energy
$U(\mathbf{r}(s_{1})-\mathbf{r}(s_{2}))$ (in units of $k_{B}T$). We will
restrict ourselves to one polymer, and similar to corresponding treatment
for flexible polymers \cite{descloizeaux} will consider the correlation
function%
\begin{equation}
G(\mathbf{k},\mathbf{k}^{\prime };N)=\left\langle \exp (-i\mathbf{kr}(N)-i%
\mathbf{k}^{\prime }\mathbf{r}(0)-\frac{1}{2}\int_{0}^{N}ds_{2}%
\int_{0}^{N}ds_{1}U(\mathbf{r}(s_{2})-\mathbf{r}(s_{1})))\right\rangle ,
\label{w40}
\end{equation}%
where the average has to be carried out in accordance with Eq.(\ref{w1}). To
derive the perturbation expansion of $G(\mathbf{k},\mathbf{k}^{\prime };N)$
we expand (\ref{w40}) in powers of the interaction energy, which is supposed
to be expanded in Fourier integral, order the integrations over the contour
length, and express all monomer positions entering (\ref{w40})$\ $through
the tangents according to $\mathbf{r}(s_{2})-\mathbf{r}(s_{1})=%
\int_{s_{1}}^{s_{2}}ds\mathbf{t}(s)$ and $\mathbf{r}(N)=\mathbf{r}%
(0)+\int_{0}^{N}ds\mathbf{t}(s)$. The integration over $\mathbf{r}(0)$ gives
the factor $(2\pi )^{3}\delta ^{(3)}(\mathbf{k}+\mathbf{k}^{\prime })$.
Similar to Sections \ref{ads-if} and \ref{ads-surf} we finally arrive at the
following expression under the average
\begin{equation}
\left\langle \exp (-i\mathbf{k}\int_{s_{2n}}^{N}ds\mathbf{t}(s))\right.
\left. \exp (-i\mathbf{Q}_{n}\int_{s_{2n-1}}^{s_{2n}}ds\mathbf{t}(s))...\exp
(-i\mathbf{Q}_{1}\int_{0}^{s_{1}}ds\mathbf{t}(s))\right\rangle ,
\label{w4xx}
\end{equation}%
where the momenta $\mathbf{Q}_{n}$, ..., $\mathbf{Q}_{1}$ are expressed by $%
\mathbf{k}$ and $\mathbf{q}_{n}$, ..., $\mathbf{q}_{1}$ using the momentum
conservation in complete analogy to flexible polymers \cite{descloizeaux}.
Using the representation of (\ref{w4xx}) through the eigenstates of the
quantum rigid rotator we obtain finally the Laplace transform of $G(\mathbf{k%
},\mathbf{k}^{\prime };N)$ as%
\begin{equation}
\int_{q_{n}}...\int_{q_{1}}U(q_{n})...U(q_{1})\left\langle 0,0|\tilde{P}(%
\mathbf{k},p)\tilde{P}(\mathbf{Q}_{n},p)...\tilde{P}(\mathbf{Q}%
_{1},p)|0,0\right\rangle .  \label{w42}
\end{equation}%
The use of the relation (\ref{m}) enables one to eliminate the propagators $%
\tilde{P}(\mathbf{Q}_{i},p)$ in favor of square matrices $\tilde{P}%
^{s}(Q_{i},p)$. Notice that to obtain from (\ref{w42}) the corresponding
expression for a flexible polymer one should replace all $\tilde{P}(q,p)$ in
(\ref{w42}) through $1/(\mathbf{q}^{2}/3+p)$, and take into account in sums
only the term with $l_{i}=0$. Thus, the comparison of (\ref{w42}) with the
corresponding expression for a flexible polymer shows that the perturbation
expansion of the correlation function $G(\mathbf{k},\mathbf{k}^{\prime };p)$
for a semiflexible polymer in powers of the interaction energy can be
represented by the same graphs as those for flexible polymers. However, the
association of the graphs with analytical expressions occurs according to
Eq.(\ref{w42}). The perturbation expansion given by Eq.(\ref{w42}) and its
straightforward generalization to many polymers is the basis for studies of
semiflexible polymers with monomer--monomer interactions.

\section{Conclusion}

\label{concl}

To conclude, we have developed the statistical-mechanical theory of
semiflexible polymers based on the connection between the Kratky-Porod model
of a semiflexible polymer and the quantum rigid rotator in an external
homogeneous field, and the treatment of the latter using the quantum
mechanical propagator method. The examples considered in this article show
that expressions and relations existing for flexible polymers can be
generalized in a straightforward way to semiflexible ones. The
correspondence is established via the replacement of the propagator of the
theory of flexible polymer through the matrix in the theory of semiflexible
polymers%
\begin{equation}
\frac{1}{k^{2}/3+p}\rightarrow (I+ikDM)^{-1}D,  \label{propag}
\end{equation}%
and consideration of an appropriate matrix element of the matrix expression
associated with the quantity under consideration. The present method
provides also a necessary framework to study problems including tangents of
polymer configurations.

\begin{acknowledgments}
The support from the Deutsche Forschungsgemeinschaft (SFB 418) is gratefully
acknowledged.
\end{acknowledgments}

\end{document}